# Ultra-Sharp Upright Photon Radiotherapy via Low Energy Extended Distance: An Alternative to FLASH for high flux Sources

Lloyd E Kamole Ghomsi[1], Clinton Gibson[1], Veng-Jean Heng[1], Ramish M Ashraf[1], Lawrie Skinner[1*]

[1]Department of Radiation Oncology, Stanford University, Stanford CA, USA

*Corresponding Author: lawrie.skinner@stanford.edu

**COI and Funding statement**

Dr. Lawrie Skinner reports current and/or recent research support from Stanford Cancer Institute (Innovation Award), and Stanford Medicine (Catalyst Award). Dr. Skinner is an inventor on intellectual property related to radiation therapy devices, including U.S. Patent No. 11,583,926 and a related U.S. patent application (US App. 16/809,427). Dr. Skinner also participates in industry-sponsored research agreements/clinical trial activities with ViewRay Systems, Inc. and Leo Cancer Care (MRAs; role: co-investigator). No specific funding was received for the present paper.

**Clinical Trial information:**

The present work does not pertain to any clinical trials.

**Acknowledgements:**

The Authors thank Dr Billy Loo For helpful discussions.

**Data sharing Statement:**

Data supporting the findings of this study are available from the corresponding author upon reasonable request. Data may be used for non-profit, academic research and educational use only. Requestors must cite the original publication and acknowledge the data source in any resulting outputs.

# Ultra-Sharp Upright Photon Radiotherapy via Low Energy Extended Distance: An Alternative to FLASH for high flux Sources

## ABSTRACT

**Purpose/Objective:** Standard 6 megavolt (MV) radiotherapy is limited by source size and secondary electron range to minimum radiological penumbra widths of ~2-3 mm. This study investigates upright radiotherapy with lower energies and extended source-to-patient distances. The objective is to double the radiotherapy beam sharpness, while maintaining the depth-dose penetration of standard setups.

**Methods and Materials:** A 2.5 MV beam from a clinical linac was delivered at a 4 m source-to-phantom distance (2.5 MV-ED). Lateral profiles and percent depth doses were measured in a solid water phantom with radiochromic film and an ion chamber. These single beam measurements were used to benchmark TOPAS Monte Carlo simulations. The validated 2.5 MV-ED model was then used to simulate upright deliveries with a conical beam geometry. These simulations were compared against equivalent plans generated for standard 6 MV-FFF coplanar deliveries at 1 m from the source.

**Results:** The 2.5 MV-ED single 28x28 mm$^2$ beam produced a measured 80%-20% penumbra of 1.0 ±0.1 mm, compared to 2.4 mm Jaw-defined penumbra for a standard 6MV-FFF beam. The doses at 10 cm depth were 52% vs. 56%, and the surface doses were 22% vs. 38% for the 2.5MV-ED and standard 6 MV-FFF respectively. Conical geometry Monte-Carlo simulations using the 2.5 MV-ED beams demonstrated significantly sharper composite dose fall-off in all cardinal directions compared to coplanar 6MV-FFF plans. For a spatially fractionated lattice example plan with 5 mm diameter high dose spheres, the 2.5 MV-ED conical approach achieved a peak-to-valley dose ratio of 4.5-5.2, compared to the 2.6-2.9 achievable with a standard 6 MV-FFF clinical system.

**Conclusion:** Low-energy, extended-distance photon beams can provide sharper penumbra and lower surface dose, while maintaining comparable depth-dose penetration as standard 6 MV setups. Combined with upright patient setups, ultra-sharp dose distributions with enhanced treatment conformity, reduced toxicity, and higher-fidelity dose modulation are possible.

# 1. Introduction

Sparing healthy tissue while conforming the high dose radiation to the target volume is the key tenet of radiotherapy. To this end modern radiotherapy linear accelerators shape dose distributions by delivering radiation from wide angular ranges with advanced intensity modulation[1,2].

The most common form of these machines is a 6 megavolt (6 MV) x-ray source at 0.8 to 1 m from the radiotherapy target volume, where MV denotes the accelerating potential. While effective, these 6 MV machines have a basic limitation: no matter how good the collimation, no matter the beam angles, minimum beam penumbra widths are limited by secondary electron ranges to ~2-3 mm (in most scenarios). These "blurry" beam edges reduce achievable coverage, contribute toxicity to neighboring organs, and limit the intricacy of target volumes. They also increase second cancer risk in surrounding tissue, which is pertinent as most second cancers after radiotherapy occur outside, but next to the target volume[3]. These limitations are becoming ever more important as patients live longer and come back for more reirradiation. Sharper beams, with less healthy tissue damage, also increase the potential for radiotherapy in non-cancer applications such as vascular disorders, cardiac arrhythmia's and neuromodulation, where side effects and long term risks are typically more of a concern than for cancer treatment[4,5].

It is well known that lower energy beams can offer sharper penumbra[6–9]. As Shown in figure 1, the secondary electron energy range, and hence the dose deposition kernels are significantly shorter for incident photon energies below 1 MeV. For Bremsstrahlung sources, the 1 MeV photon energy roughly corresponds to the average energy of a 3 MV beam. Planning studies with 2-3 MV energy beams have been shown to be superior for superficial targets[10,11]. These lower energy beams, however, also have poorer penetration, making them generally inferior to 6 MV with standard setups for deeper targets [10,11].

An alternative radiotherapy approach, that is gaining renewed interest is to keep the linac source still and instead rotate the patient while in an upright position[12,13]. For particle ion beam radiotherapy (proton, carbon etc.), this removal of the gantry has clear size and cost savings[14]. While the cost and space savings from eliminating the gantry are smaller for photons than for particles, upright geometry still offers meaningful advantages. It creates an opportunity to integrate novel, more powerful photon sources, incorporate advanced imaging capabilities, and allow greater flexibility in source-to-patient distance [13,15].

In this work, novel upright patient setups with extended-distance collimation and lower energy beams are investigated with a view to creating ultra-sharp dose distributions. There are four main drivers for this approach: (i) Sharper penumbra, (ii) improved penetration in the patient, (iii) lower room shielding requirements. (iv) efficient delivery of non-coplanar, conical arcs.

## 2. Materials & Methods

To achieve sharper (smaller) beam penumbra the major contributing factors should be considered. These are (i) collimator leakage & scatter (ii) patient scatter and (iii) geometric penumbra. Collimator leakage can be solved with increased thickness. Collimator scatter is also small; typically collimator scatter-to-primary ratios are less than 10% [16]. Patient scatter and geometric penumbra, however, can be improved and are considered in detail in sections 2.1 and 2.2.

### *2.1 Patient scatter*

Patient scatter consists of both photon scattering and secondary electron scattering. Photon scattering in the patient is mainly Compton scattering. Although Compton scatter occurs slightly more frequently at wider angles as energy reduces, the cross sections are relatively constant with energy in the MV region (figure 1B). Little can be done about Compton scattered photons in the patient.

Secondary electron scattering, however, can be dramatically reduced in range by reducing the incident photon energy, as shown in figure 1 A. Essentially, incident photon energies of 1 MeV or less produce secondary electrons, which travel 1 mm or less in water (on average).

Specifically, this information, shown in figure 1, was obtained from taking the Kelin-Nishina formula for differential cross sections[17] and the Compton scattering formula that determines the secondary electron energy as a function of angle after Compton interactions[18]. This function of energy versus angle was then converted into range in water using standard tables[19]. By weighting the normalized differential cross section at each angle (figure 1B) and then summing over all the angles the average range of a Compton scattered electron is obtained as a function of incident photon energy (figure 1A). While this is not the complete physics, more than 90% of the interactions in this 0.2-3 MeV photon energy range are Compton scattering[9]. These average electron ranges also compare closely to calculated dose spread arrays, see figure 1A inset (taken from [20,21]). Both the average ranges and the 2D arrays indicate dose spread of ~3mm at 6MV and less than 1 mm at 3 MV and lower energies.

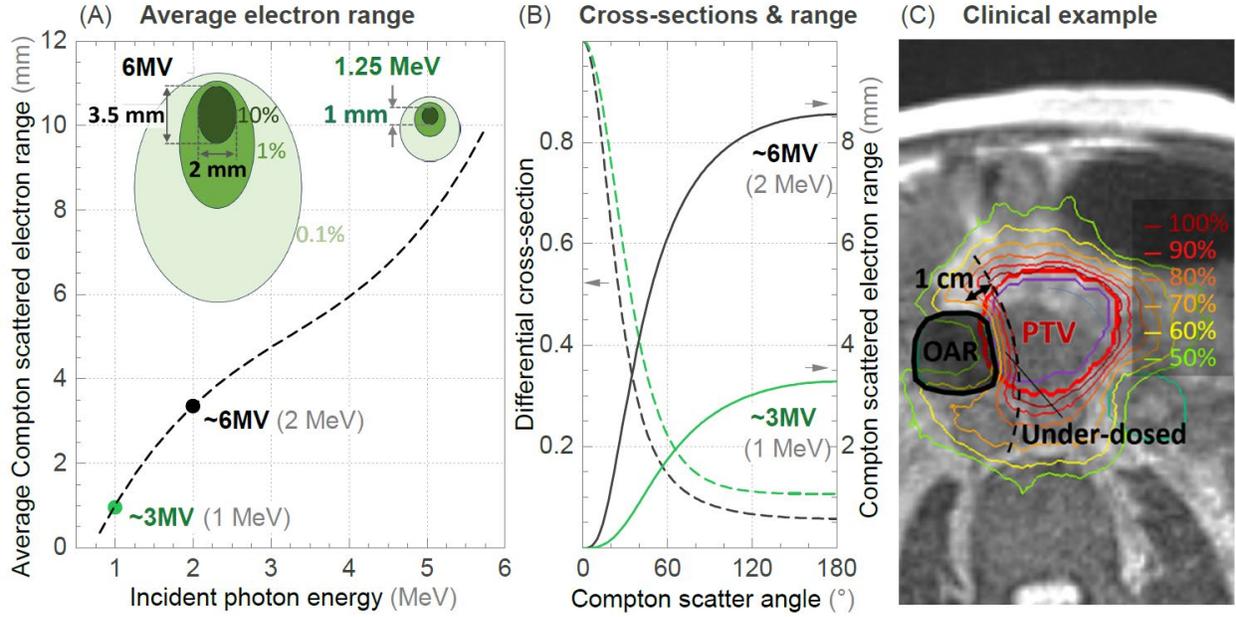

**Figure 1.** (A) The average Compton scattered electron range as a function of incident photon energy (dashed line). Inset are dose spread arrays for 1.25 MeV and 6MV incident photons in water, which agree well with the calculated average range. Note that at 6MV the approximately 3 mm electron range limits the sharpness of the dose deposition. (B) Shows the standard differential cross sections (dashed lines) and Compton scattered electron range functions (solid lines), from NIST databases, that were combined to obtain the curve in (A). The 1 MeV energy which corresponds to the average energy of a 3MV beam is given in green. The 2MeV energy which corresponds to the average energy of a 6MV beam is given in black. (C) A clinical example, from the ViewRay MRIdian 6MV-FFF system with double-stacked double-focused MLCs, yet the dose gradients were still insufficient to cover the target and spare the adjacent organ at risk.

### 2.2 *Geometric penumbra & source size at extended distance*

Geometric penumbra ($P_{geo}$), which arises from the source having a non-zero size also adds to penumbra width. Using small angle approximations, the formula simplifies to

$$P_{geo} \cong \frac{\Delta s}{SAD/D_{coll} - 1} \qquad \text{Eqn. 1}$$

where $D_{coll}$ is the collimator-target distance, where the target can be approximated as the center point of the Planning Target Volume (see figure 2B). Since patient clearance requires $D_{coll}$ to be approximately 40 cm or greater, source size ($\Delta s$) and the source-axis distance (SAD) are the main determinants of the geometric penumbra. As shown in figure 2A, the smaller geometric penumbra at extended SAD distances allows for larger sources. This small geometric penumbra, means that more fluence can be achieved with the same power-density on the Bremsstrahlung target, easing cooling requirements. To more fully consider the relative dose rates of different size sources, energies, and distances (SADs), the following formula can be used (equation 2)

$$DR \cong DR_{std} \cdot \frac{SAD_{std}^2}{SAD^2} \cdot \frac{\Delta s^2}{\Delta s_{std}^2} \cdot \frac{\eta}{\eta_{std}} \qquad \text{Eqn. 2}$$

Where DR is the dose rate from a source at a given SAD, with a source size Δs, and Bremsstrahlung efficiency, η. Taking a standard 6MV source ($DR_{std}$) with 1 m SAD, and a 1 mm Δs. For a source of the same material, the ratio $\eta/\eta_{std}$ can be estimated as $E/E_{std}$. Then, scaling a standard source with a 1mm source size at 1 m SAD to a 4 m SAD and a 4 mm source size. The dose rates would be identical, but the extended source would have a smaller geometric penumbra (≈0.57 mm vs ≈1 mm). Going down from 6MV to 2.5 MV would bring dose rates down from 1400 cGy/min to ~585 cGy/MU. This is without increasing the power-density on the tungsten target above that of current clinical linacs.

Hence, a larger source size can largely compensate for the dose-rate loss from the extended SAD distance of the source, yet the extended distance brings sharper geometric penumbra and slower depth-dose fall-off in the patient (i.e. improved penetration).

### *2.3 Depth-Dose Fall-off*

Since the dose rates and geometric penumbra can be reasonably achieved for large sources at extended distances, the improved depth-dose fall-off of this extended distance setup can be leveraged. As shown in Figure 2 C, the extended distance confers a reduced inverse square fall-off at depths past the maximum dose ($d_{max}$). Specifically, the extended distance means that past $d_{max}$ the dose reduces more slowly with increasing depth, reducing the ratio of shallow dose near $d_{max}$ and the prescribed dose at the target depth. Ultimately this means less monitor units are needed to achieve a given prescribed dose when compared to the same beam quality at a standard 1 m source distance. Another way to think of this is that the extended beams have ~4x lower divergence than standard setups, at the same field size.

To further improve the depth-dose fall-off of 2.5 MV beams, a beam hardening strategy using high-z filtration was also investigated. Lead filters, chosen for their high electron number, strongly attenuate photon energies below 1 MeV, while allowing a high transmission fraction of the >1 MeV photons (figure 2D). This technique becomes more favorable at the 1-3 MeV beam energies as high-z filters are more effective below 1 MeV, where photo electric absorption is strong (figure 2E).

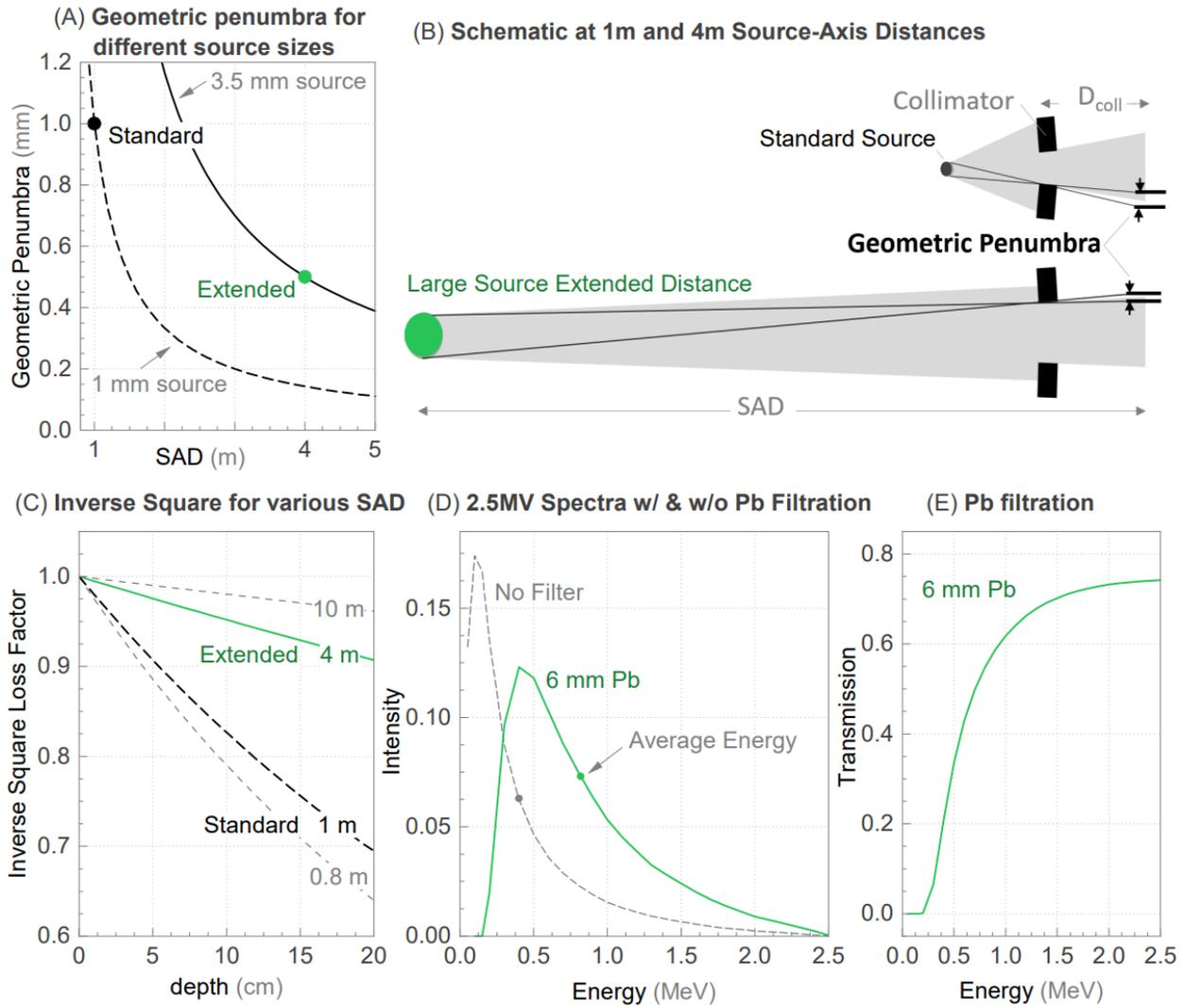

**Figure 2.** (A) Since Geometric penumbra reduces with extended distance setups (green circle), larger source sizes (solid line) can be used with extended distance setups. (B) Schematic showing how the extended distance geometry, with long source to axis distance (SAD), reduces geometric penumbra. (C) Extended distance setups (solid line) have improved depth-dose fall-off compared to standard setups (dashed black line). (D, E) Lead filtration is effective at attenuating photon energies below 1 MeV. The solid line in (D) shows the spectrum used in the Monte-Carlo simulations (figure 4) that matched the measured depth-dose and beam profiles. The circles in (D) indicate the mean energies with and without lead filtration. The no-filter spectrum for 2.5 MV (dashed line in D, was digitized from [22]).

## 2.4 Low-Energy Extended-Distance Setup

Informed by the basic physical phenomena investigated in S2.1-S2.3, a low-energy extended-distance experimental setup with a source to phantom distance of 4 m using a 2.5 MV-FFF beam from a Varian TrueBeam was chosen (Varian Medical systems, Palo Alto, CA, USA). To filter out most of the <0.2 MeV photons, a 6 mm thick lead filter was placed immediately after the mylar window of the linac head (figure 2E). This filtered, extended distance beam will be denoted as 2.5 MV-ED. Measurements were made with this setup at a 28x28 mm² field size, which was formed by a collimator made of copper-tungsten alloy bars that were 6 cm thick (75wt% tungsten and 25wt% copper). This collimator was placed with the upstream edge 3.5 m from the source (0.5 m from the phantom, see figure 4C). The custom collimator was manually aligned to the beam using the light field and a digital level. The uncertainty from this manual alignment was approximately 0.5 mm in the plane transverse to the beam, and 1-2° in the vertical and left-right directions.

EBT3 radiochromic (Ashland, Bridgewater, NJ, USA) Film was placed inside a 30x30x15 cm³ solid water phantom transverse to the beam at a depth of 3cm. Films were scanned using an Epson Expression 1000XL scanner, at 300 dpi, in transmission scan mode. Films were analyzed using the FIJI distribution of the ImageJ software [23] using the red channel, and following the procedures detailed in [24]. Specifically, the OD was obtained by taking the ratios to regions of interest with no film, then OD of an unirradiated film from the same sheet was used to obtain background subtraction. Calibration films were obtained from 6 MV exposures at dose levels from 0 to 2 Gy in 0.25 Gy increments. A Varian Truebeam was used for calibration at 100 cm SSD and the depth of maximum dose (1.5 cm). The linac output was measured immediately before irradiation of the calibration films using the AAPM TG51 protocol [25]. The film calibration data was applied in ImageJ using the Rodbard (NIH image). For the line profiles (figure 4A) a line 24 pixels wide was averaged (approximately 2 mm wide).

Depth-dose measurements were also made with an Exradin A10 parallel plate ion chamber with a 5.4 mm collector diameter, 2 mm gap, and <0.004 cm²/g buildup (Standard Imaging, Middleton, WI, USA). Multiple chamber depths were measured while keeping the phantom surface-to-source distance constant. A PTW Unidos webline electrometer was used with a -300V bias setting. The effective point of measurement was taken to be the front window of the chamber.

The measured beam profiles and depth-doses were used to benchmark TOPAS Monte-Carlo simulations that mirrored the experimental setup [26]. This model consisted of the same material and final collimator geometry as the physical setup, but the source and primary collimator were simplified to a gaussian with 1 mm positional spread in X and Y and a cutoff at 1.2mm in X, Y. The angular spread was also set to 1 ° with a 1.5 ° cutoff. The spectrum is that given in figure 2D which was obtained through several trials with varying lead filter attenuation to match the measured depth-dose curves (see figures 2,4). While the simulated source size and shape likely do not precisely match the true source size and shape, the extended distance setup means that effects of source size are de-magnified, by a factor ~4x, allowing for good agreement with experimental data, without the need to fine tune the source shape. The dose scoring grid used a cubic 0.5 mm³ voxel size. Multiple runs were performed each with approximately $10^9$ histories. The Monte-Carlo parameter files are given in the supplemental information. Once the single beam correspondence was established, further models with multiple beams simulating conical arc setups were performed.

## 2.5 Upright Conical Arc Radiotherapy

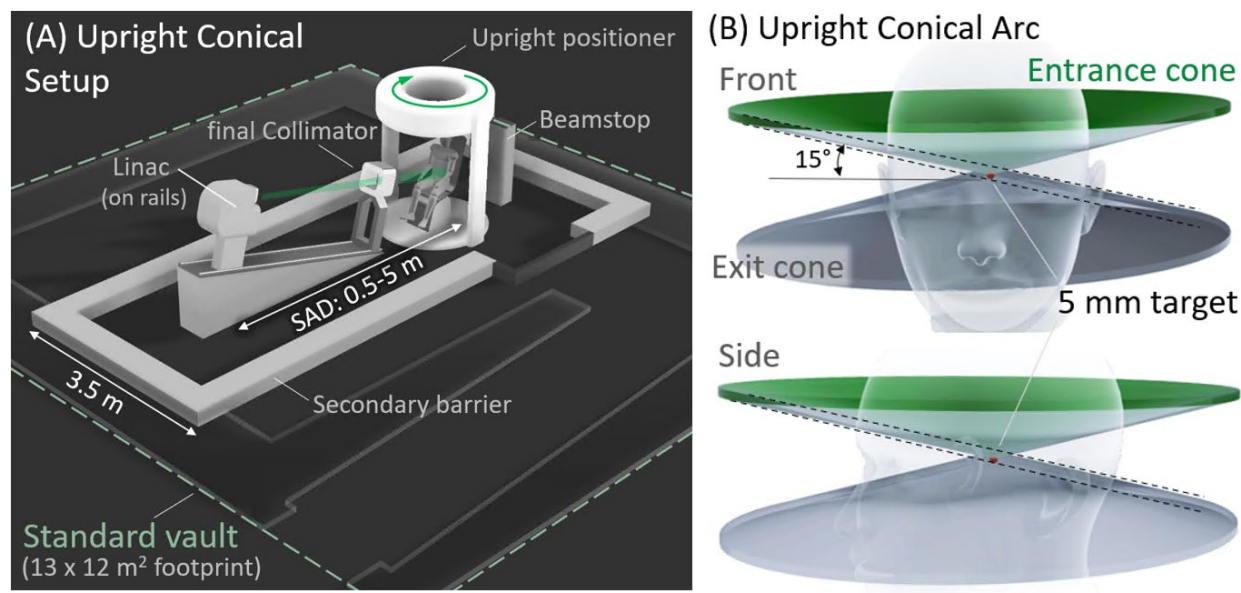

**Figure 3.** (A) Upright setup example shown relative to a standard photon radiotherapy vault. Even accounting for the extended distance, the constant beam angle simplifies primary beam shielding and reduces the system footprint by at least a factor 2, compared to the 18MV vault example from NCRP151. The limited maximum energy ~6MV also reduces the secondary barrier thickness requirements as essentially no photon-neutrons are produced. (B) By tilting the beam axis relative to the patient rotation axis, dose deposition is transformed from a flat disc into a double-cone. Specifically, entrance and exit doses then physically diverge and no longer add up at the same anatomic locations.

While the 2.5 MV-ED setup is designed to provide sharp transverse beam profiles, it does not help dose gradients parallel to the beam path. For this, a common strategy in megavoltage x-ray radiotherapy is to deliver non-coplanar beams. Non-coplanar radiotherapy has been shown to improve dose distributions for a variety of treatment sites and is particularly useful in SBRT and SRS settings where organs at risk border the treatment target[1,27–31]. In horizontal patient setups, non-coplanar delivery requires either couch rotations, multiple sources, or a multi-axis source positioning system. Clinically, couch rotations are used sparingly on C-arm linacs due to collision risks and imaging limitations[32]. Whereas machines that employ multi-axis source positioning systems such as the CyberKnife and Zap-X machines, are often limited to smaller lower power sources or shorter source-to-patient distances [33–35].

Angling the fixed beam off-horizontal, with an upright rotating patient setup, enables delivery of non-coplanar arcs that form a double-cone dose distribution (see figure 3). Entry and exit doses then no longer sum on the same plane and diverge on the superior-inferior axis. Hence, this conical setup offers fast, simple delivery of arcs, where the radiation is increasingly "defocused" in the vertical plane with increasing distance from the target volume. Note that to form conical arcs, this beam angle tilt needs to be with respect to the axis of rotation of the rotational platform, it is different to tilting the patient from vertical, or delivering supine arc therapy with a couch kick, which do not form conical dose distributions.

Conical arcs were simulated in TOPAS v3.6 using the 2.5 MV-ED beam described in S2.4, calculated on a 28 cm diameter cylinder of water, which was rotated around the patient rotation axis, with a 15° offset between the beam angle and the patient rotation axis (as shown in figure 3). Partial pseudo-arcs were approximated by two opposing sets of 9 beams calculated at 8° intervals from -32° to +32° (see figure 5). Comparison was made to a coplanar delivery of 6MV-FFF beams from a Varian TrueBeam on the same phantom, with a standard SAD of 1 m, and jaw defined fields with 6 mm openings and the same 8° spacing and 64° wide fan, normalized to 100% covers 99% of a 4 mm diameter 4 mm long cylindrical target volume. The choice of 5 mm jaw defined fields was to provide optimal sharpness and coverage for the standard system. See supplemental information for parameter files. Coplanar standard delivery was chosen for comparison to the conical upright plans because the approximate delivery time and complexity between the two setups are comparable (3D imaging at the treatment position, fast delivery time). Comparing to non-coplanar supine geometries, for example using couch rotations on a Varian truebeam, greatly increases delivery complexity and limits 3D imaging, which is not the case for the conical upright delivery. Alternatively comparing to a CyberKnife would mean comparing a delivery time of approximately 5-10x longer, than a coplanar C-arm linac delivery.

## *2.6 Spatially fractionated radiotherapy*

Using the extended distance, conical arc beam geometry with the same 15° angle offset between the beam and patient rotation axes, dose was calculated for a pair of 5 mm wide jaw openings, and four patient positions (i.e., four conical arcs) to form a 2x2x2 array of 5 mm dose targets on a body centered cubic lattice, with centers separated by 20 mm. Each pseudo-arc consisted of beams 45 beams equally spaced 8° apart. The Monte Carlo calculations were compared to an equivalent clinical protocol using VMAT (volumetric Modulated Arc Therapy) on a Varian Truebeam with an HD-MLC, which was beam matched to Varian golden beam data, and using the 6X-FFF mean energy (calculated dose to water, on a water phantom Using Acuros External beam v 15.6.05). This comparison was chosen as it is a common clinical standard with a delivery speed that is comparable to the 4-conical arcs of the simulated upright setup. In addition to the set of eight 5 mm diameter spheres another 8 "valley dose" sphere volumes, and a pair of ring structures (2cm and 3cm expanded from the PTV) were used in the optimization (Varian photon optimizer v15.6.05). The dicom files for this coplanar comparison plan are provided in the supplemental information.

The Monte Carlo uses the same 0.5mm$^3$ voxel size from the dose scoring grid as the calculations in sections 2.4 and 2.5. The Peak and Valley values were obtained from the averaging the sets of 8 minimum and maximum point doses around the lattice locations.

# 3. Results

## 3.1 Single beam profiles and depth-dose measurements

Figure 4 shows the measured beam profiles and depth-dose distributions for the single beam, 2.5 MV extended-distance (2.5 MV-ED) setup described in section 2.4. The 80-20% penumbra width of this 2.5 MV-ED beam was $1.0 \pm 0.1$ mm compared to $2.4 \pm 0.1$ mm for the standard 6MV-FFF beam (which was a jaw-defined beam with a 1 m source-axis distance). The dose at 10 cm deep ($PDD_{10}$) was similar between the 2.5 MV-ED and standard 6MV-FFF beams: 52% vs 56 % respectively. Note that larger field sizes would also yield commensurately higher $PDD_{10}$ for both energies. The surface dose of the 2.5 MV-ED beam was measured to be $22\% \pm 1\%$ with a depth of maximum dose of 6 mm $\pm$ 1 mm, compared to $38\% \pm 1\%$ and $13 \pm 1$ mm for the 6MV-FFF beam. Monte-Carlo measurements (dotted lines in figure 4) agree within the uncertainties of the measured profile and depth dose curves.

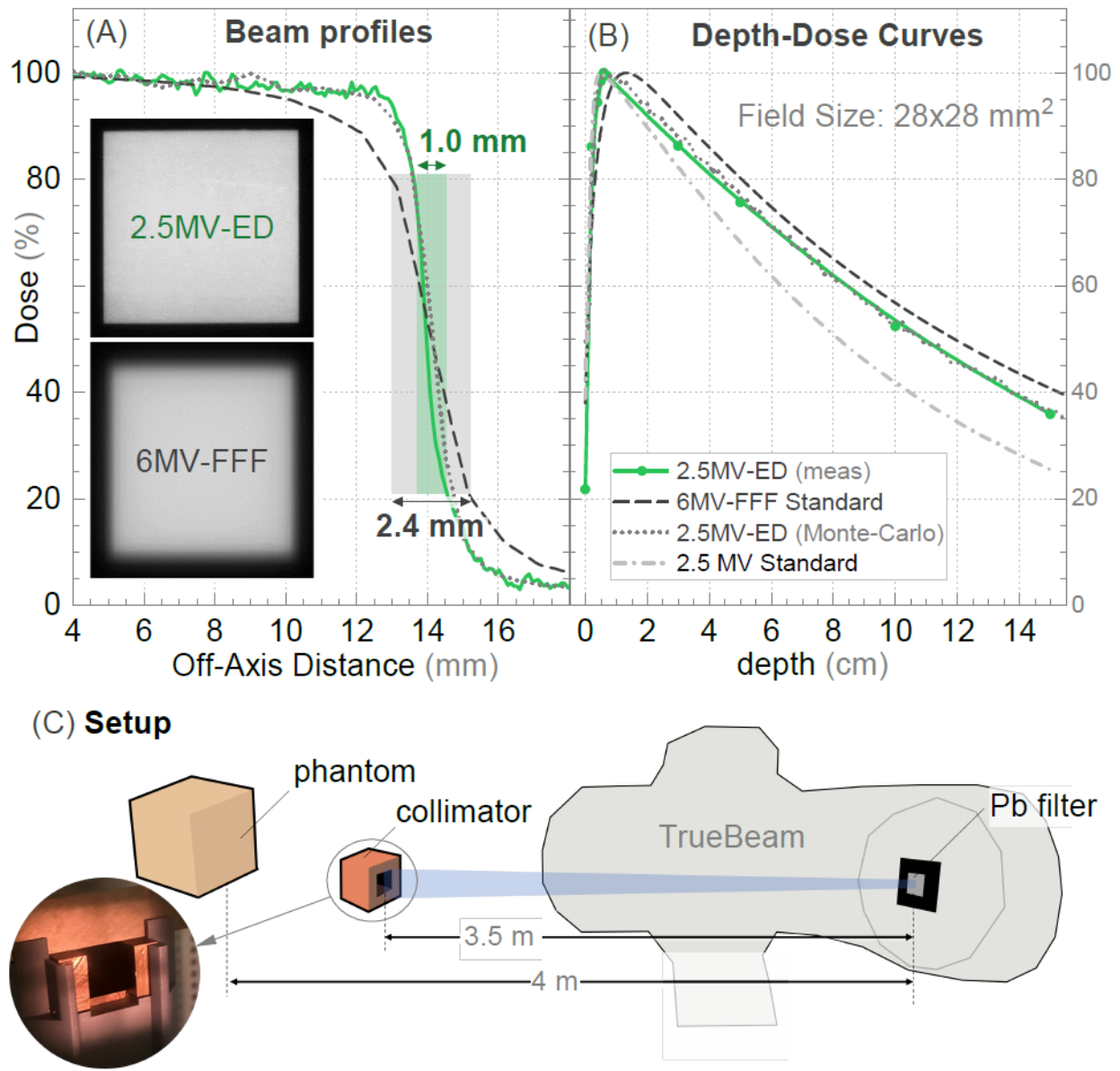

**Figure 4.** The 2.5 MV-ED (extended distance) beam is significantly sharper than standard 6MV beam, while the depth-dose is comparable to the standard 6MV beam. (A) Beam profiles at 3cm depth in a solid water phantom. The solid line is the 2.5 MV Extended-distance beam which had an SAD of 4m, measured with radio chromic film. The grey dotted line is the same profile from Monte-Carlo calculations. Also shown is the calculated profile for a standard 6MV-FFF beam from a Varian TrueBeam at 1m SAD (dashed grey line). Inset top is the measured film from which the green line in (A) was obtained. Inset bottom is the calculated dose plane of the standard 6MV-FFF beam (see S2.4 for detail). (B) Are the same beams in the depth direction in a 30x30x20 cm³ solid water phantom. The only difference in data type is that the green data (solid line and circles) in (B) were measured with an A10 parallel plate ion chamber at the green circular points shown (see S2.4 for detail). The light do-dashed line is the measure PDD curve for the unfiltered 2.5 MV beam at a standard 100 cm source to surface distance. (C) is a setup diagram for the measured data (solid green lines in A, B), including a photograph of the custom Cu-W collimator.

## 3.2 Upright Conical Arc dose distributions

Figure 5 compares the results for the upright conical arc and standard coplanar setups described in section 2.5 normalized to the same 100% coverage of the 4 mm diameter target. Profiles through the center of the target show faster dose fall off in all three cardinal directions for the conical 2.5 MV-ED delivery. In the Sup-Inf and Ant-post directions, 80-20 penumbra was 1.0 mm (± 0.25 mm) for the 2.5 MV-ED vs >2 mm for the standard delivery. Note that the dose normalization is higher in the standard 6 MV-FFF setup to achieve equal target coverage, which results in 132% hotspot vs 113% in the 2.5 MV-ED. In the left-right direction 50% dose was reached at 8.8 mm vs 12.5 mm for the standard beam.

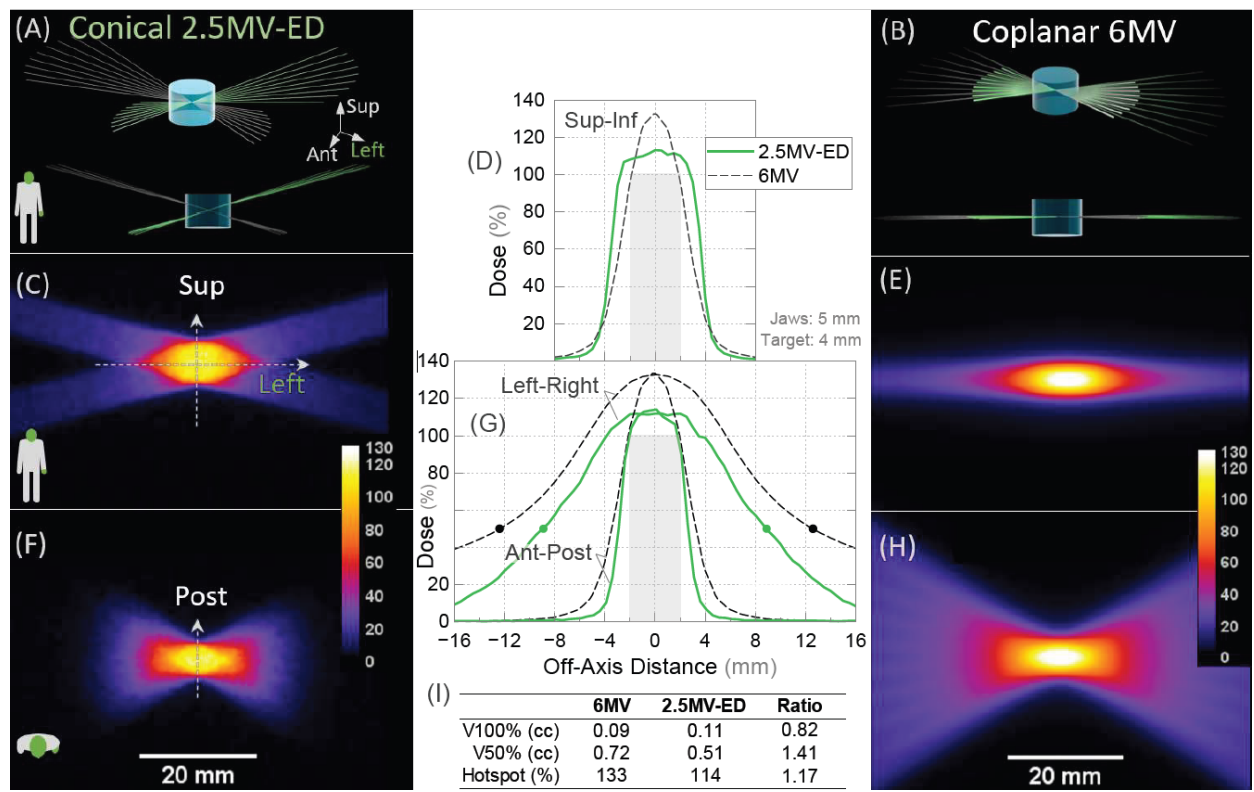

**Figure 5.** Dose distributions for a 5 mm wide jaw setting normalized to cover 100% of a 4 mm diameter, 4 mm long cylindrcial target. (A) and (B) show the beam arrangementsfor the extended-distance, 2.5 MV-ED conical geometry and the standard distance coplanar geometry. In both cases two opposing fans of beams are used, with 8 degree spacing between beams and a 64 degree-wide fan. In the conical setup the beams are all angles 15 degrees off of the axial direction. (C) & (E) show coronal slices through the target center. (F) and (H) are matching slices in the axial plane. (D) and (G) are the dose profiles through the center of the target in each of the cardinal directions. Green solid lines correspond to the 2.5 MV-ED conical geomtery and the black dashed lines are the 6 MV-FFF coplanar results. The circles in (G) indicate the 50% dose level. The Metrics in (I) confirm that for this 5 mm jaw setting the 2.5 MV-ED provides, larger 100% volume coverage, and faster dose fall-off. The over-coverage in the Sup-Inf direction is because of the sharp penumbra and the jaw being 1 mm wider than the target volume. The conical geometry has the cost that additional low dose is deposited in regions sup and inf of the target.

## 3.3 Spatially fractionated Radiotherapy

The results of the spatially fractionated calculations (methods section 2.6) are given in figure 6. Specifically, 2.5 MV-ED and the standard (Varian TrueBeam with HD-MLC) are compared. Figure 6 (A)-(C) show the setup for the 2 x 2 x 2 high dose target array, while (D)-(I) show dose profiles and dose distributions.

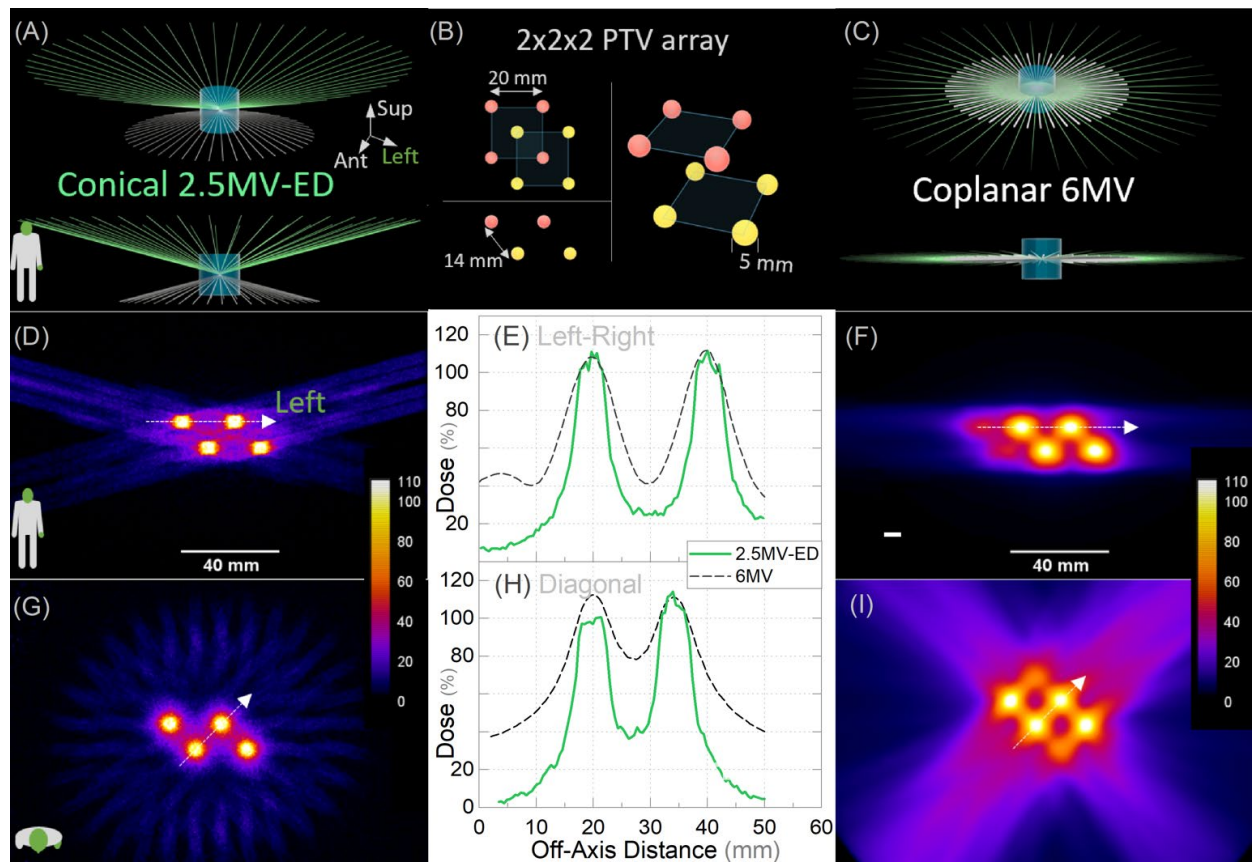

**Figure 5**. Spatially fractionated radiotherapy example. (A) and (C) show the beam arrangements for the extended-distance, 2.5 MV conical geometry and the standard distance coplanar geometry. In both cases a full circle of beams are used, (the 2.5MV beams ore open conformal fields with an 8-degree spacing, the 6MV beams are Volumetric Modulated Arc Therapy (VMAT) arcs, calculated on a Clinical Truebeam with an HD-MLC. (B) is the high dose target volumes for these lattice plans. It consists of eight 5 mm diameter spheres. In the conical setup the beams are all angles 15 degrees off of the axial direction. (D) & (F) show coronal slices through the target. (G) and (I) are matching slices in the axial plane. (E) and (F) are the dose profiles through the center of the target in the Left-right and diagonal (right-anterior to left-posterior). Green solid lines correspond to the 2.5 MV-ED conical geometry and the black dashed lines are the 6MV coplanar results. The 2.5 MV-ED provides faster dose fall-off despite the lack of intensity modulation, achieving higher peak-to-valley dose ratios, that the clinical system cannot achieve at this small target size and separation.

## 3.4 Alternative modes of an Extended distance upright Photon source.

Table 1 gives several example setups calculated using equations 1 and 2 as well as using the obtained measured results at the 4 m SAD. Specifically, table 1 takes the dose rate of a clinical system (Varian TrueBeam) as the standard reference (STD). The dose rates and penumbra widths are then estimated for different SAD source sizes, and energies, for the same power density of electrons incident on the target. While full Monte carlo dosimetric calculations of the various source sizes, and distances is beyond the scope of the present paper, it is still instructive to consider the effects on the major parameters. Which is mainly that source size can be used to increase the dose rate at extended distance, with a smaller than might be expected deleterious effect on beam penumbra. It should be clearly stated that the main technology required to achieve this is 20-40x higher than standard electron beam current. Fundamentally, the extended distance allows us to select a smaller solid angle radiation cone, which therefore creates lower divergence beams for a given field size. The three non-standard example modes given are (i) SHARP mode – this uses lower energy and extended distance to improve the penumbra and optimize penetration. (ii) FAST & LARGE mode – which uses standard energy (~6MV) at extended distance with a large source size to creates the same geometric penumbra as the STD setup, with improved penetration, dose rate and maximum field size. This FAST & LARGE mode would have ~3x higher dose rate and slower PDD fall-off than a STD setup for the same radiologic penumbra. (iii) FLASH mode – this is the source size, energy, and distance combination needed to achieve FLASH dose rates. This was obtained from taking 8MV and 0.6 m SAD as minimum values from which source size was adjusted to achieve FLASH dose rates. The FLASH setup numbers are intended as comparison. A dedicated multisource FLASH device is expected to achieve improved metrics [36,37]. Even so, a single large source setup like this may be useful for simple FLASH treatment sites such as superficial targets and whole brain.

| Mode | Energy | Source size ($\Delta s$, mm) | Source-Axis Distance | Inverse square Factor $\frac{SAD_{std}^2}{SAD^2}$ | Source size Factor $\frac{\Delta s^2}{\Delta s_{std}^2}$ | Brem. efficiency $\frac{\eta}{\eta_{std}}$ | Geometric Penumbra (mm) | **Dose rate (cGy/min)** | **Radiologic 80-20 penumbra (mm)** |
|---|---|---|---|---|---|---|---|---|---|
| STD. | 6 MV | 1 mm | 1 m | 1 | 1 | 1 | 1.00 | **1400** | **2.4** |
| SHARP | 2.5 MV | 4 mm | 4 m | 1/16 | 16 | 0.417 | 0.57 | **585** | **1.0-1.2** |
| FAST & LARGE | 6 MV | 7 mm | 4 m | 1/16 | 49 | 1 | 1.00 | **4290** | **2.4** |
| FLASH | 8 MV | 7 mm | 0.6 m | 2.78 | 49 | 1.333 | ~10.0 | **~240,000** | **>10** |

*Table 1 Dose rates at isocenter and at the depth of maximum dose in a water phantom for different setups. Extended distance 3.5mm source compared to current clinical standards assuming the same power density of the electron beam on the target. The dose rate quoted is at isocenter and at the pedth of maximum dose.*

## 4. Discussion

The 2.5 MV-ED beams measured here achieved more than double the "sharpness" of standard 6 MV-FFF radiotherapy setups, while also closely matching the depth-dose fall-off of the standard setup (figure 4). When going to lower energy one immediate concern is skin toxicity; this was a key driver for radiotherapy originally moving to higher energies. The measured skin dose for the 2.5 MV-ED beams was 22%, which is lower than standard 6 MV-FFF beams, because of the extended air path, results in less electron contamination reaching the patient (or phantom). Since the depth-dose fall-off of the 2.5 MV-ED beam is also comparable to 6 MV-FFF beams, the skin dose in 2.5 MV-ED treatment plans is expected to be the same or lower than for standard 6 MV-FFF setups.

The conical arc plans calculated on a 28 cm wide cylindrical phantom (figure 5) further confirm that the 2.5 MV-ED beams can produce sharp composite plans. The dose-distributions show penumbra widths of ~1 mm in both the sup-inf and ant-post directions through the center of the target, which is significantly sharper than the same plan in the standard 6 MV-FFF coplanar calculations (dashed lines, figure 5). The L-R direction, which is closer to the beam directions, show 50% dose reduction in 8 mm, which is also faster fall-off than the coplanar 6 MV-FFF system. In this direction the increased fall off is a result of the 15 deg non-coplanar beams, whereas in the 6 MV-FFF coplanar example the entry and exit beam doses add up in most locations.

The conical-arc upright setups investigated here provide non-coplanar beams without additional complexity, collision risk, imaging limitations, and negligible additional delivery time compared to coplanar arcs. Although this conical geometry represents an intermediate between full solid angle coverage and simple coplanar beams, human bodies are approximately cylindrical. Hence the majority of the useful beam angles, outside the head, are typically close to the axial direction.

A drawback to the conical arc approach compared with coplanar arcs is the slight increase in photon path length and a higher volume of tissue receiving a low dose. A 15° beam tilt is approximately an additional 4% in photon path length, which is most likely an acceptable increase in integral dose.

A relevant clinical system for comparison is the 3 MV Zap-X linac. This is a specialized brain radiosurgery system, with a short source to axis distance of 45 cm and the ability to deliver beams from a wide range of solid angles. Due largely to its low energy, the ZAP-X system has been shown to provide class-leading dosimetric fall-off and superior clinical plan quality for small field sizes [8,34,38,39]. Although the ZAP-X system has sharp penumbra in the range 1-2 mm for small field sizes, at field sizes above 1cm, that system's penumbra width increases to 2-3 mm, consistent with the ~1.5-2 mm source size and short source-to-collimator distance[8]. The ZAP-X approach of using lower energy and short source-target distances is useful in the brain, which is 2% of the human body. The rapid depth-dose fall-off and limited field size of that short SSD approach largely prohibits its application to most non-brain, non skin treatment sites. In contrast, the longer source-collimator and source-patient distances of the present work enable improved penetration (higher percent depth dose), and is capable of large field sizes, while maintaining sharp beam penumbra, opening up all human body treatment sites.

One specific area where efficient delivery of sharper beams may be of use is spatially fractionated radiotherapy. Figure 6 shows that significant dosimetric improvement over clinical 6MV systems is possible with just a few arcs in the extended-distance conical geometry. Peak-Valley dose ratios of 4.5-5.2 were obtained for the 2.5 MV-ED setup, whereas the clinical 6MV-FFF system yielded only 2.6-2.9 (see figure 6). This is despite having intensity modulation which the 2.5 MV-ED calculations lacked.

A clinical radiotherapy system in the energy range 2 to 8 MV with an extendable source distance, and upright patient positioning, also brings multiple secondary benefits unavailable to current paradigms. Specifically, the lower energy fixed beam simplifies shielding barriers and system footprint: A beam stop can replace the primary shielding barrier, and the lack of fast neutron radiation allows for denser lead and steel shielding instead of concrete. As shown in figure 3, an upright system with the capability to extend up to ~5 m source-patient distances, would have a footprint less than half that of a standard radiotherapy linac. Such extended-distance upright setups also bring more efficient whole-body intensity modulated treatments. Currently these typically require five or more isocenters, on a C-arm linac, or long beam-on times (~30mins) on a tomotherapy linac [40].

As shown in table 1, a linac with a large source designed to give useful dose rates at extended distances (~4 m) would also be capable of higher dose rates at short patient to source distances and or larger source sizes. Since extended distances make both the geometric penumbra smaller and the PDD fall-off slower, there is a net benefit in using larger sources at extended distance vs small sources at standard distances. Fast treatments may be used to eliminate respiratory motion between imaging and treatment. Ultimately, a mixture of high-dose rate and ultra-sharp modes, may also prove to be optimal.

One alternative perspective on this low-energy extended distance approach is that it can provide a complementary mode of use to a FLASH-capable photon radiotherapy machine (where FLASH is radiotherapy requires dose rates exceeding 40 Gy/s). Since FLASH will probably not be optimal for all treatment sites, an ultra-sharp mode would provide more utility to a FLASH-capable photon machine. Further inspection of table 1 and FLASH literature[36,37,41], indicate that FLASH dose rates will most likely require larger photon beam penumbra, approximately ~5 mm or more, as a combination of shorter source distances, larger sources, or higher energies are needed to achieve FLASH dose rates[36,37,41]. Specifically, the large penumbra widths of photon FLASH mean that neighboring organs at risk will likely receive higher physical doses than conventional deliveries. This may be OK where the FLASH sparing effects are sufficiently high, but the sparing effect will depend on the specific anatomy, dose rate at that location, and tissue type. Hence, there will likely be many clinical scenarios where using "ultra-sharp" modes to reduce dose to neighboring organs, is more useful than FLASH (and vice versa), and a dual mode machine will be a more effective use of clinical resources.

One aspect that sharper beams require is tighter patient immobilization and positioning. As deviations arising from patient motion or beam alignment may reduce the composite beam sharpness. Modern motion management with 2D and 3D x-ray imaging, along with surface guidance, are typically at or above this mm-level precision required.

Further work in this area of sharper radiotherapy machines includes investigating a range of energies, cone-angle offsets, and distances toward the above goals. A full treatment planning system and optimizer for these beam energies and geometries would also provide more clinical insight into the comparative benefits of these lower energy, extended distance setups.

## 5. Conclusion

Many aspects of conventional 6 MV photon radiotherapy are now approaching practical limits in terms of achievable dose gradient and conformity. By using extended distances, 6MV-like beam penetration can be maintained at 2-3 times lower maximum energies. The secondary electron range is then 2-3 times shorter, hence sharper lateral dose fall-off is possible. Additionally, conical geometry beams offer a simple approach to fast-delivery of non-coplanar arc radiotherapy in the upright position. These sharper dose distributions can be used to improve organ sparing, or alternatively increase hypofractionation, as well as provide finer spatially fractionated radiotherapy.

This work is pertinent to upright photon radiotherapy machines, as most of the technological pieces needed already exist. With lower energy extended distance beams, there is opportunity for upright photon radiotherapy to go beyond being low cost and instead provide increased precision and reduced toxicity.

**Declaration of generative AI and AI-assisted technologies in the manuscript preparation process**

During the preparation of this work the author(s) used Open AI GPT 5.2 and Anthropic Claude 4.5 Sonnet, to assist in refining the title and abstract. These were tools also used to provide suggestions for improvement of the main text. No AI was used in the figures or figure data. After using this tool/service, the authors fully reviewed and edited the content as needed and take full responsibility for the content of the published article.

Cancer Radiotherapy. *Sci. Rep.* **2025**, *15* (1), 17604. https://doi.org/10.1038/s41598-025-02150-4.